\documentclass[12pt]{article}
\usepackage[utf8]{inputenc}
\usepackage{amsmath}
\usepackage{tensor}

\usepackage{amsfonts}
\usepackage{IEEEtrantools}

\usepackage{graphicx}
\DeclareGraphicsExtensions{.pdf,.png,.jpg, .gif}
\begin{document}

\baselineskip=18pt

\setcounter{footnote}{0}

\setcounter{figure}{0}
\setcounter{table}{0}

\begin{titlepage}

{\begin{flushright}
 {\bf      }
\end{flushright}}

\begin{center}
\vspace{1cm}

{\Large \bf  AdS$_3$ and AdS$_2$ Magnetic Brane Solutions}

\vspace{0.8cm}

{\bf Ahmed Almuhairi}

\vspace{.5cm}

{\it  Department of Physics, University of California, \\ Santa Barbara, California
93106, USA}

\end{center}
\vspace{1cm}

\begin{abstract}

We investigate AdS$_3$ and AdS$_2$ magnetic brane solutions within the consistent truncation of AdS$_5 \times S^5$ supergravity.  The AdS$_3$ solution extends earlier work by allowing a general embedding of the magnetic $U(1)$ in $SO(6)$.  We determine the ratio of strong- and weak-coupling entropies as a function of this embedding. Further, by considering crossed magnetic fields in different $U(1)$'s we are able to construct a solution that runs from AdS$_5$ in the UV to AdS$_2 \times R^3$ in the IR. We find the notable result that there is a zero temperature entropy at strong coupling but not at weak coupling.  We also show that the AdS$_2$ solution and at least some of the AdS$_3$ solutions are stable within the truncation.

\end{abstract}

\bigskip
\bigskip

\end{titlepage}

\section{Introduction}

The AdS/CFT correspondence introduced by Maldacena \cite{Maldacena:1997re} has helped shed light on behaviors of systems otherwise far too complicated to tackle. Specifically, it has given us a method by which strongly interacting regimes of theories can be studied quite easily through considering the dual gravity theory which is mostly described by geometry. Also, it allows us to study responses to the field theory to background electromagnetic fields by introducing gauge fields in the bulk, which is another simplifying feature.

Many AdS/CFT systems with a background magnetic field have been studied, especially in the case of the AdS$_4$ magnetic brane used to investigate 2+1 gauge theories in magnetic fields \cite{Hartnoll:2007ai, Hartnoll:2007ih, Hartnoll:2007ip}. Also, AdS$_5$ case has been tackled in \cite{D'Hoker:2009mm}, which was motivated by RHIC physics which involves strongly coupled gauge theories in magnetic fields \cite{Son:2009tf}. A slight generalization of the configuration of the magnetic fields can prove to be useful in shedding light on more exotic features of the system.

Another interesting aspect of AdS/CFT is the correspondence in 2 dimensions, i.e. AdS$_{2}$/CFT$_{1}$.  As opposed to common belief that smaller dimension means simpler physics, this case of the duality has proven to be the least understood \cite{Strominger:1998yg}. This is partly due to the disconnected feature of the boundary of the AdS$_2$ space, giving us two disconnected dual descriptions commonly understood to be systems described by conformal quantum mechanics. Recent interest stems from the emergence of quantum critical behavior in this duality. Such behavior, involving non-Fermi liquid behavior, was studied in \cite{Faulkner:2009wj, Faulkner:2010tq}. A new setup showing the the emergence of the AdS$_2$ space-time will prove useful to the cause of understanding the duality in 2 dimensions.

Driven by these motivations we consider bulk solutions that interpolate between AdS$_5$ in the UV regime and AdS$_3$$\times$T$^2$ in the IR along with the presence of a magnetic field. A similar system was studied in \cite{D'Hoker:2009mm} but we generalize the problem by considering a magnetic field that is comprised of a linear combination of three $U(1)$'s of $SO(6)$. In this generalization, the linear combination is parameterized by elements of the unimodular tensor, $T_{i j}$, that is in the 20' representation of $SO(6)$ which emerges from the deformations of the 5-sphere in the AdS$_5$$\times S^5$ after truncation \cite{Cvetic:2000nc}. We consider a simple diagonal form for the tensor with three parameters $T_{1,2,3}$. At finite temperature, the large coupling limit is the product of a BTZ black hole and a T$^2$. The entropy is found to scale as $T$, temperature, just as in \cite{D'Hoker:2009mm}. The added feature in our result is that the entropy is also a function of the linear combination parameters, $T_i$. 
 
We also consider the effect of this generalization on the ${\cal{N}} = 4$ SYM theory. The charged particles are now charged under a linear combination of the $U(1)_R$ magnetic fields. It's entropy too depends on the linear combination parameters. The relation between the entropies takes the form $S_G = F(T_1, T_2) S_{{\cal{N}} = 4}$ where the peak of the function $F(T_1, T_2)$, occurring at $T_i = 1$, corresponds to the result $S_G = \sqrt{\frac{4}{3}} S_{{\cal{N}} = 4}$ obtained in \cite{D'Hoker:2009mm}. We also find solutions for arbitrary $F(T_1,T_2)$ below the peak value. This is an indication of the influence of the chosen linear combination on the physics of the problem.

Next, we look at the other system of interest which interpolates between AdS$_5$ and AdS$_2$$\times$T$^3$. The magnetic field considered has the added feature that each $U(1)$ points in a different direction in physical space; one along each direction of the torus. We find that this system can be solved exactly, and we obtain the interpolating solution for all $r$. We find that the solution interpolates between AdS$_5$ at large $r$ and the product of an extremal two dimensional black hole with T$^3$. The black hole thus has finite entropy at zero temperature hinting at a 1 dimensional dual field theory with a degenerate ground state. 

Across the duality, we show that AdS$_2$ magnetic field configuration does not alter the temperature scaling of the entropy. We then show that we still have the regular free ${\cal{N}} = 4$ SYM theory, with zero entropy at zero temperature.

Finally, we investigate the stability properties of the solutions at hand, as this is often a problem for condensed matter applications of AdS/CFT duality.  We consider small perturbations of the unimodular tensor away from the identity. We find two classes of solutions having different $m^2$. We find that for both space-times the $m^2$'s are well above the respective $m_{BF}^2$. Thus both the space-times are stable.
 
In this paper, we start in section 2 by constructing the space-time that interpolates between AdS$_5$ and AdS$_3\times$T$^2$ and we compute the relevant entropy. In section 3 we discuss the ${\cal{N}} = 4$ SYM dual field theory and compare the entropies. Then we go on to investigate the thermodynamics of the AdS$_5$/AdS$_2\times$T$^2$ space-time in section 4. In section 5 we present an argument showing that the entropy of the free field limit of ${\cal{N}} = 4$ SYM still scales as temperature for the AdS$_2$ magnetic field configuration.  In section 6 we discuss the stability of the space-times considered by computing the $m^2$ of the unimodular tensor fields.

\section{AdS$_5$/AdS$_3\times$T$^2$ Gravity Theory}

We use the results obtained by \cite{Cvetic:2000nc} in truncating the 10 dimensional  IIB theory to a 5 dimensional Einstein-Maxwell theory. The Kaluza-Klein reduction Ansatz is given by
\begin{equation}
d\hat{s}^2_{10} = \Delta^{1/2}ds_5^2 + g^{-2}\Delta^{-1/2} T^{-1}_{ij}D\mu^i D\mu^j
\end{equation}
where
\begin{align}
\Delta &\equiv T_{i j} \mu^i \mu^j, \ \ \ \mu^i \mu^i = 1 \\
D\mu^i &\equiv d\mu^i + g A^{ij}_{(1)} \mu^j
\end{align}
where $T_{i j}$ is a symmetric 6$\times$6 unimodular tensor used to represent the 20 scalars in the 20' representation of $SO(6)$. This also represents the L $=2$ deformation of the 5-sphere. The $A^{i j}$ are the 1-form potentials, antisymmetric in $i$ and $j$, that represent the 15 $SO(6)$ Yang-Mills gauge fields. The radius of the five compact dimensions is given by the inverse of $g$. Also just to define some notation, here are a few important relations
\begin{align}
D T_{i j} &\equiv d T_{i j} + g A^{i k}_{(1)} T_{k j} + g A^{j k} _{(1)} T_{i k} \\
D B^{i j}_{(p)}  &= d B^{i j}_{(p)} + g A^{i k}_{(1)} \wedge B^{k j}_{(p)} + g A^{j k}_{(1)} \wedge B^{i k}_{(p)} \\
 D_{\mu} B^{\mu}_{i j} &= \nabla_{\mu} B_{i j}^{\mu} + g A^{i k}_{\mu} B^{\mu}_{k j} + g A^{j k}_{\mu} B^{\mu}_{i k} \\
 D_{\mu} T_{i j} &= \partial_{\mu} T_{i j} + g A^{i k}_{\mu} T_{k j} + g A^{j k}_{\mu} T_{i k}
\end{align}
where $\nabla$ is the usual covariant derivative. The first two equations are in form notation while the third is in terms of tensors. \\
\indent The truncated lagrangian has the form
\begin{align}
{\cal{L}}_5 &= R \ast {\bold{1}} - \frac{1}{4} T_{i j}^{-1} \ast D T_{j k} \wedge T_{k l}^{-1} D T_{l i} - \frac{1}{4} T_{i k}^{-1} T_{j l}^{-1} \ast F_{(2)}^{i j} \wedge F_{(2)}^{k l} - \tilde{V} \ast{\bold{1}}\\
&= \left( R - \frac{1}{4}T_{i j}^{-1} D_{\mu} T_{j k}  \ T_{k l}^{-1} D^{\mu} T_{l i}    - \frac{1}{8} T^{-1}_{i k} T^{-1}_{j l} F^{i j}_{\mu \nu} F^{\mu \nu}_{k l}  -  \tilde{V}                    \right) \ \ast {\bold{1}}
\end{align}
Where $ \tilde{V}$ is a potential given by
\begin{equation}
\tilde{V}= \frac{1}{2} g^2 (2 T_{kl} T_{kl} - ( T_{kk})^2)
\end{equation}
Also we have omitted Chern-Simons terms that will not contribute to the physical system of interest to us.
The unimodular tensor field equation that one obtains from this lagrangian is
\begin{align}
-D_{\mu}(T_{i k}^{-1}  D^{\mu} T_{k j}) &= -2g^2(2 T_{i k} T_{j k} - T_{i j} T_{k k}) +\frac{1}{2} T^{-1}_{i k} T^{-1}_{l m} F^{l k}_{\mu \nu} F^{\mu \nu}_{m j}  \nonumber\\
& - \frac{1}{6} \delta_{i j} [-2g^2(2 T_{k l} T_{k l} - (T_{k k})^2) +\frac{1}{2} T^{-1}_{p k} T^{-1}_{l m} F^{l k}_{\mu \nu} F^{\mu \nu}_{m p}  ]  \label{tmotion}
\end{align}
\indent The solutions we are looking for are those that interpolate between AdS$_5$ at high energies and AdS$_3$$\times$T$^2$ and low energies. We are also interested in incorporating a magnetic field that is tangent to the boundary directions. We also assume Lorentz invariance in the boundary spatial directions orthogonal to the magnetic field. The metric Ansatz then takes the form
\begin{equation}
ds_5^2 =  - U(r) dt^2 + \frac{dr^2}{U(r)} + e^{2 V(r)}((dx^1)^2+(dx^2)^2) + e^{2 W(r)}dy^2
\end{equation}
We are interested in a magnetic field that is a linear combination of three different $U(1)$'s taken out from the group $SO(6)$. We can obtain this via the following choice of $T_{i j}$ and the Maxwell stress tensor $F_{(2)}^{i j}$
 \begin{equation}
T_{ij} =
\begin{bmatrix}
T_1 & 0 & 0 & 0 & 0 & 0 \\
0 & T_1 & 0 & 0 & 0 & 0 \\
0 & 0 & T_2 & 0 & 0 & 0 \\
0 & 0 & 0 & T_2 & 0 & 0 \\
0 & 0 & 0 & 0 & T_3 & 0 \\
0 & 0 & 0 & 0 & 0 & T_3
\end{bmatrix}, \
F^{ij}_{(2)} =
\begin{bmatrix}
0 & -\lambda_1 & 0 & 0 & 0 & 0 \\
\lambda_1 & 0 & 0 & 0 & 0 & 0 \\
0 & 0 & 0 & -\lambda_2 & 0 & 0 \\
0 & 0 & \lambda_2 & 0 & 0 & 0 \\
0 & 0 & 0 & 0 & 0 & -\lambda_3 \\
0 & 0 & 0 & 0 & \lambda_3 & 0
\end{bmatrix}
    {\cal F}_{(2)}
\end{equation} 
where
\begin{equation}
{\cal F}_{(2)} = {\cal B} dx^1 \wedge dx^2
\end{equation}
and the unimodular condition becomes $T_1 T_2 T_3 = 1$. With our choice of $T_{i j}$ and Maxwell tensor  $F_{(2)}^{i j}$, we go ahead to solve the tensor field equations (\ref{tmotion}). We do this with the assumption that $T_{i j}$ is a constant in the near horizon regime. Due to complexity of solving for the $T_i(\lambda_i)$ we instead solve for $\lambda_i(T_i)$ to get the relations
\begin{equation}
\frac{\lambda^2_1}{T^2_1} = \frac{\lambda^2_2}{T^2_2} + \frac{-8 g^2}{{\cal F}_{(2)}^2} T_3(T_1 - T_2)  = \frac{\lambda^2_3}{T^2_3} + \frac{-8 g^2}{{\cal F}_{(2)}^2} T_2(T_1 - T_3)
\end{equation}
plus the cyclic permutations in 1, 2, and 3.
Next, the langrangian for the truncated theory becomes
\begin{equation}
  {\cal{L}}_5 = R  - \frac{1}{4} \left(  \sum^{3}_{i = 1} \frac{\lambda_i^2}{T_i^2}  \right) {\cal F}_{(2)}^2 - \tilde{V}
  \end{equation}
 where the $\tilde{V}$ is now
 \begin{equation}
\tilde{V} = -4g^2(\frac{1}{T_1} + \frac{1}{T_2}+ \frac{1}{T_3}) \label{potential}
\end{equation}
Implementing our Ansatz for the metric, we obtain the following Einstein's equations
  \begin{align}
  rr:& \ \ U' V' + \frac{1}{2} U' W' +\frac{1}{2} U'' + 2 U V'^2+ U W'^2 + 2 U V'' + U W'' = 4 \tilde{L}^{-2} + \frac{2}{3} e^{-4V} B^2\\
  11:&\ \ U' V' + 2 U V'^2 + U V' W' + U V'' = 4 \tilde{L}^{-2} - \frac{4}{3}e^{-4V} B^2 \\
  yy:&\ \ U' W' + 2U V' W' + U W'^2 + U W'' = 4 \tilde{L}^{-2} + \frac{2}{3}e^{-4V} B^2 \\
  tt:&\ \ 2 U' V' + U' W' + U'' = 8 \tilde{L}^{-2} + \frac{4}{3}e^{-4V} B^2
  \end{align}
where
  \begin{align}
  \tilde{L}^{-2} &= \frac{g^2}{3}(\frac{1}{T_1} + \frac{1}{T_2}+ \frac{1}{T_3}) \\
  B^2 &= \frac{1}{4} \left(  \sum^{3}_{i = 1} \frac{\lambda_i^2}{T_i^2}  \right) {\cal B}^2
  \end{align}
 We now look  for the required solutions. For $B = 0$, we have the solution representing AdS$_5$ given by the usual metric
 \begin{equation}
  ds_5^2 =  \frac{r^2}{\tilde{L}^2}  \left(  - dt^2+(dx^1)^2+(dx^2)^2+dy^2\right)+ \frac{\tilde{L}^2}{r^2}dr^2
  \end{equation}
  This is the expected UV solution, located at $r \rightarrow \infty$, of our theory. In this region the magnetic field becomes negligible and can be ignored. The next case is with $B \neq 0$; this will represent the field theory at longer length scales, and thus is expected to be located at finite $r$. The field equations produce the following solution at the IR
\begin{equation}
  ds_5^2 =  - \frac{3(r^2 - r^2_+)}{\tilde{L}^2} dt^2 + \frac{\tilde{L}^2 dr^2}{3(r^2 - r^2_+)} + \frac{B\tilde{L}}{\sqrt{3}}((dx^1)^2+(dx^2)^2) + \frac{3r^2}{\tilde{L}^2}dy^2
\end{equation}
This IR solution represents the product between a BTZ black hole and a torus, T$^2$. We see that at zero temperature we obtain AdS${_3}$$\times$T$^2$.

\indent The entropy of the BTZ black hole is given by $S = \frac{\pi}{3} c T L_y$, where $T$ is temperature, $L_y$ is the length of the $y$ direction, and $c$ is the central charge. We obtain the central charge using the Brown-Henneaux formula \cite{Brown:1986nw}, $c = 3 l/2 G_3$, where $l$ is the AdS$_3$ radius given by $l = \tilde{L}/\sqrt{3}$, and $G_3$ is the 3 dimensional gravitational constant. We can write $G_3$ as $\sqrt{3} G_5/\tilde{L} B V_2$, where $V_2$ is the volume of the two compact dimensions $x^{1,2}$. Simplifying the expression further, we write $G_5$ in terms of relevant string theory constants to get
\begin{align}
c_G &= \frac{4 \pi^3 g^{-5} \tilde{L}^2 B V_2  V_5 16 \pi}{15 (2 \pi)^7 g_s^2 l_s^8} \\
&= \frac{2^7 \pi^5 \tilde{L}^2 g^3 B V_2 V_5 N^2}{15 {\cal{G}}^2}
\end{align}
where $V_5 =$ Unit Volume of the $\mu_i$ dimensions $/ (2 \pi)^3$, and ${\cal{G}}$ comes from defining the integral of the 5-form as $\int \hat{*} \hat{G}_{(5)} \equiv g^{4} {\cal{G}}$, and $N$ is the number of branes, which is given by
\begin{equation}
N = \frac{1}{16 \pi G_{10}  T_3}\int \hat{*} \hat{G}_{(5)}
\end{equation}
We note that $B$ is a function of the charges, and thus a function of the elements of the unimodular tensor, $T_{i j}$. Now that we have the central charge on the gravity side, we proceed to obtain the central charge across the duality and compare.

\section{AdS$_3\times$T$^2$ Dual ${\cal N} = 4$ Super Yang-Mills Field Theory}

We go on now to calculate the entropy of the free field limit of ${\cal{N}} = 4$ SYM theory in an external magnetic field. We consider the ${\cal{N}} = 4$ theory in ${\cal{N}} = 1$ terms, with a $U(1)$ R-Symmetry. In this theory, we have $N^2$ complex scalars of each charge, $\lambda_1$, $\lambda_2$, and $\lambda_3$. These are the scalars in the chiral multiplet. We also have $N^2$ vector fields of charge 0, these are the gauge fields. From the charges of the complex scalars we obtain the charges of the fermions of the chiral multiplet and the Gauginos to be
\begin{align}
\beta_1 &= \frac{\lambda_1 - \lambda_2 - \lambda_3}{2} \\
\beta_2 &= \frac{\lambda_2 - \lambda_3 - \lambda_1}{2} \\
\beta_3 &= \frac{\lambda_3 - \lambda_1 - \lambda_2}{2} \\
\alpha &= \frac{\lambda_1 + \lambda_2 + \lambda_3}{2}
\end{align}
So we have $N^2$ Weyl spinors of each charge $\beta_1$, $\beta_2$, and $\beta_3$. These are the fermions of the chiral multiplet. And finally we have $N^2$ Weyl spinors of charge $\alpha$, these are the Gauginos. \\
Next we write down the logarithm of the partition function for each type of field. For the details of obtaining these we refer the reader to \cite{D'Hoker:2009mm}. The expressions are
\begin{align}
&\bullet \mathrm{Complex \ Scalars}:  \nonumber \\ 
& \ \ \ \ \ln Z_{\phi}(q_{\phi}) = -2 \frac{g |q_{\phi} {\cal{B}}| V_2}{2 \pi} \sum_{n = 0}^{\infty} \frac{L_y}{2 \pi} \int_{-\infty}^{\infty} dp_y \ln\left( 1 - e^{-\beta \sqrt{p_y^2 + (2n+1)g |q_{\phi} {\cal{B}}|}}    \right) \\
&\bullet \mathrm{Weyl \ Spinors}: \nonumber  \\
& \ \ \ \  \ln Z_{\psi}(q_{\psi}) = \frac{g |q_{\psi} {\cal{B}}|V_2}{2 \pi} \sum_{n=0}^{\infty} \sum_{\alpha = \pm 1} \frac{L_y}{2 \pi} \int_{-\infty}^{\infty} dp_y \ln \left(1+e^{-\beta \sqrt{p_y^2 + g |q_{\psi} {\cal{B}}|(2n + 1 - \alpha)}}  \right)\\
&\bullet \mathrm{Gauge \ Fields}:  \nonumber \\
&\ \ \ \ \ln Z_V = -2 \frac{V_2 L_y}{(2 \pi)^3} \int d^3p \ln \left( 1 - e^{- \beta |p|}\right)
\end{align}

\begin{figure}[!t]
  \hfill
  \begin{minipage}[t]{.45\textwidth}
    \begin{center}  
     \includegraphics[width=7cm]{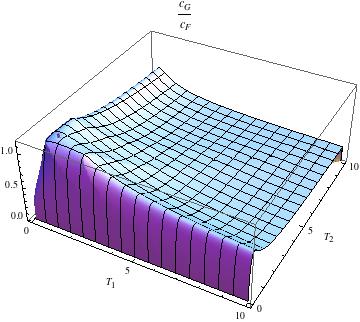}  
    \end{center}
  \end{minipage}
  \hfill
  \begin{minipage}[t]{.45\textwidth}
    \begin{center}  
     \includegraphics[width=7cm]{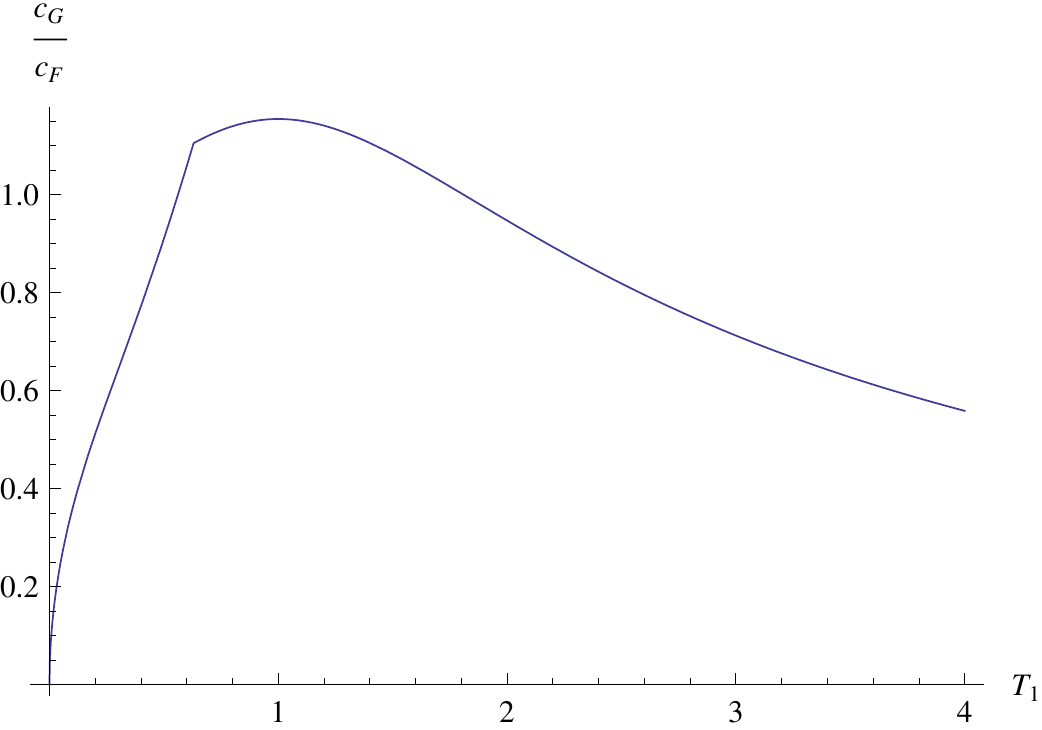}   
    \end{center}
  \end{minipage}
  \caption{ These are plots of the ratio between the central charges of the AdS$_3$$\times$T$^2$ theory and the ${\cal{N}} = 4$ SYM free field theory. On the right we have the ratio as a function of $T_{1,2}$. On the left we have a slice through $T_1 = T_2$. \label{CC}}
  \hfill
\end{figure}

We are interested in calculating the entropy in the extremely low temperature limit. We use the standard formula, $S = (1-\beta \frac{\partial}{\partial \beta} ) \ln Z$. In the required limit, the only contribution we have is the fermionic term with $n = 0,\ \alpha = 1$ which gives
\begin{align}
S &= \frac{\pi}{3} c_F L_y T, 
\\ c_F &= \sum_{\psi} \frac{1}{2} \frac{g |q_{\psi} {\cal{B}}| V_2}{2 \pi} \\
&= \frac{1}{2}(|\alpha| + |\beta_1|+ |\beta_2|+ |\beta_3|)\frac{g |{\cal{B}}|V_2}{2 \pi}N^2\end{align}
We compare the central charges of the two theories across the duality. We consider the ratio $\frac{c_G}{c_F}$. Simplifying this, we find it to be a function of any two of the set ($T^1$, $T^2$, $T^3$).  Figure (\ref{CC}) shows a plot of $\frac{c_G}{c_F}$ as a function of $T^1$ and $T^2$, along with a slice through the $T_1 = T_2$ plane.

The result of \cite{D'Hoker:2009mm} is reproduced at the point $T_1 = T_2 = 1$, which corresponds to the peak of $\frac{c_G}{c_F}$ where $S_G = \sqrt{\frac{4}{3}} \ S_F$. We also note the wide range of values of $\frac{c_G}{c_F}$ as we vary the linear combination of our gauge fields in group space.

\section{AdS$_5$/AdS$_2\times$T$^3$ Gravity Theory}

In this section we search for a different type of solution motivated by finding a field theory dual that lives in one dimension. We thus look for a solution to the truncated IIB theory that interpolates between AdS$_5$ at large $r$ and AdS$_2 \times$T$^3$ at small $r$. The Ansatz we choose has both translational and rotational symmetry in the toroidal directions. This is manifest in the following metric
\begin{equation}
ds_5^2 =  - U(r) dt^2 + \frac{dr^2}{U(r)} + e^{2 V(r)}\left((dx^1)^2+(dx^2)^2 +(dx^3)^2\right)
\end{equation}
The magnetic field we consider in this section is a linear combination of fields that point in different directions in actual space; one along each toroidal direction. Also, each magnetic field will be chosen to be part of a different $U(1)$ as before. While the unimodular tensor, $T_{i j}$, remains the same, the Maxwell tensor will then be chosen as
\begin{equation}
F^{ij}_{(2)} =
    \begin{bmatrix}
    \begin{bmatrix}
    0 & -\lambda_1 \\
    \lambda_1 & 0
    \end{bmatrix}
    {\cal F}_{(2)}^1  \\ \\
    & \begin{bmatrix}
    & -\lambda_2 \\
    \lambda_2 & 0
    \end{bmatrix}
    {\cal F}_{(2)}^2  \\ \\
    & & \begin{bmatrix}
    0 & -\lambda_3 \\
    \lambda_3 & 0
    \end{bmatrix}
    {\cal F}_{(2)}^3
    \end{bmatrix}
    \end{equation}
where we set the magnetic fields to point in orthogonal directions, as shown below
\begin{align}
{\cal F}_{(2)}^1 &= {\cal B}_1 \ dx^2 \wedge dx^3 \\
{\cal F}_{(2)}^2 &= {\cal B}_2 \ dx^3 \wedge dx^1 \\
{\cal F}_{(2)}^3 &= {\cal B}_3 \ dx^1 \wedge dx^2
\end{align}
Upon solving the Tensor field equations for $\lambda_i(T_i)$ we find
\begin{equation}
\frac{\lambda^2_1}{T^2_1} ({\cal F}_{(2)}^1)^2 = \frac{\lambda^2_2}{T^2_2} ({\cal F}_{(2)}^2)^2 -8 g^2 T_3(T_1 - T_2)  = \frac{\lambda^2_3}{T^2_3}({\cal F}_{(2)}^3)^2 -8 g^2 T_2(T_1 - T_3) \label{ads2tmotion}
\end{equation}
plus the cyclic permutation in 1, 2, and 3.
The lagrangian is slightly modified from the previous case
\begin{equation}
  {\cal{L}}_5 = R  - \frac{1}{4} \left(  \sum^{3}_{i = 1} \frac{\lambda_i^2}{T_i^2} ({\cal F}_{(2)}^i)^2  \right)  - \tilde{V}
  \end{equation}
  where $\tilde{V}$ is given in eq. (\ref{potential}). Varying the action we obtain the following Einstein's field equations  
  \begin{align}
  rr:& \ \ \frac{3}{2}U' V' +\frac{1}{2} U'' + 3 U V'^2+ 3 U V''  = 4 \tilde{L}^{-2} + \frac{2}{3} e^{-4V} \vec{B}^2\\
  11:&\ \ U' V' + 3 U V'^2 + U V'' = 4 \tilde{L}^{-2} + \frac{2}{3} e^{-4V} \vec{B}^2 - 2e^{-4V} (B_2^2 + B_3^2) \\
  22:&\ \ U' V' + 3 U V'^2 + U V'' = 4 \tilde{L}^{-2} + \frac{2}{3} e^{-4V} \vec{B}^2 - 2e^{-4V} (B_3^2 + B_1^2) \\
  33:&\ \ U' V' + 3 U V'^2 + U V'' = 4 \tilde{L}^{-2} + \frac{2}{3} e^{-4V} \vec{B}^2 - 2e^{-4V} (B_1^2 + B_2^2) \\
  tt:&\ \ \frac{3}{2} U' V' + \frac{1}{2} U'' = 4 \tilde{L}^{-2} + \frac{2}{3}e^{-4V} \vec{B}^2
  \end{align}
  with
  \begin{align}
\vec{B}^2 &= B_1^2 + B_2^2 + B_3^2 \\
B_i^2 &= \frac{1}{4}\frac{\lambda_i^2}{T_i^2} ({\cal B}_i)^2\\
\tilde{L}^{-2} &= \frac{g^2}{3}(\frac{1}{T_1} + \frac{1}{T_2}+ \frac{1}{T_3})
\end{align}
 These field equations, along with the scalar field equation result for $\lambda_i$, \ref{ads2tmotion}, constrain our choice of the magnitudes of the magnetic fields, charges, and the $T_{i j}$ entries. We are forced to set
 \begin{align}
T_1& = T_2 = T_3 = 1 \\
\frac{\lambda_1^2}{T_1^2} ({\cal B}_1)^2 &= \frac{\lambda_2^2}{T_2^2} ({\cal B}_2)^2 = \frac{\lambda_3^2}{T_3^2} ({\cal B}_3)^2 \equiv \lambda^2 {\cal B}^2 \equiv 4 B^2 \\
\tilde{L}^{-2} &= g^2
\end{align}
The field equations then take a simpler form and can be solved exactly. At the UV limit we impose $B = 0$. The result, as expected, is the usual AdS$_5$ solution given by
\begin{equation}
ds_5^2 =  g^2r^2  \left(  - dt^2+(dx^1)^2+(dx^2)^2+(dx^3)^2\right)+ \frac{ dr^2}{g^2r^2}
\end{equation}
Next, we turn on the magnetic field and solve the field equations exactly for $U(r)$ and $V(r)$. We note that there are two types of solutions to the Einstein's field equation; One where $V(r) = V_0$ is taken to be a constant, and the other is where $V(r)$ varies with $r$. We expect that the constant $V_0$ solution be a limiting case of the varying solution. We choose to place the constant value at $r_+$ which will be the location of the horizon of the black hole. Solving for $U(r)$ we find that it does indeed have a solution that vanishes at an $r_+$ giving us a horizon. We also find that when we place the $V_0$ at $r_+$ the solution of the black hole becomes an extremal one. It's metric is given by
\begin{equation}
ds_5^2 =  - 8g^2(r - r_+)^2 dt^2 + \frac{ dr^2}{8g^2(r - r_+)^2} + \frac{B}{g\sqrt{2}}\left((dx^1)^2+(dx^2)^2+(dx^3)^2\right)
\end{equation} 
which is the product between a two-dimensional black hole and a three torus, T$^3$ as required. Calculating the temperature of this black hole we find it to be $T \sim (r - r_+) = 0$ as expected of a black hole at extremality. The interesting aspect of this black hole is that we can still go ahead and calculate an entropy at zero temperature. We do so by using the standard formula, $S = \frac{A}{4 G_5}$, with the area, $A$, being
\begin{equation}
A = \frac{B^{3/2}V_3}{g^{3/2} 2^{3/4}}
\end{equation}
As for the gravitational constant, we have the usual AdS/CFT relation, $G_5 = \frac{\pi}{2 N^2}$. We note that the relations used here are much simpler than those used in in section 2 because of the simpler form of $T_{i j} = \delta_{i j}$. This means that the $S_5$ is not deformed which simplifies many of the equations in the truncation process. Finally we compute the entropy to be
\begin{equation}
S = \frac{B^{3/2}V_3 N^2}{g^{3/2} 2^{7/4} \pi}
\end{equation}
We emphasize again that this entropy is finite at zero temperature. This implies that our gravity theory is dual to a field theory with a degenerate ground state at large coupling. We proceed now to evaluate the entropy of the free field limit of the dual theory.

\section{AdS$_2\times$T$^3$ Dual ${\cal N} = 4$ Super Yang-Mills Field Theory}

The particle content here is identical to the one in section 3. The difference comes in the value of the assigned charges. The complex scalars each are assigned the charge $\lambda$. Three of the Weyl fermions are assigned the charge $-\lambda/2$ and the Gaugino is assigned the charge $\frac{3}{2} \lambda$. The claim now is that not much has changed from the previous case in section 3. Since the scalars transform under the $\bold{6}$ of $SO(6)$ each complex scalar component will be charged under only one gauge field, thus its energy eigenstates will not change and the partition function will not be altered. Since the gauge field is not charged, we expect that it also will attain the same partition function. As for the Weyl fermions, they transform as a $\bold{4} $ of $SO(6)$. Thus each component will be charged under a linear combination of the different $U(1)$'s and hence of the different magnetic fields pointing in the different directions. However, we will show with an example that this does not alter the answer except by a change in the magnitude of the magnetic field it feels. Here is a simple heuristic argument: Since the fermions are charged equally under all three fields, then it will feel a resultant field equal to the square root of the sum of each magnetic field squared, ${\cal{B}}_R = \sqrt{ {\cal{B}}^2 + {\cal{B}}^2 + {\cal{B}}^2} = \sqrt{3} {\cal{B}}$. As for the direction of ${\cal{B}}_R$, this depends on the details of $U(1)$ linear combination for each fermion. \\
Here is the example: consider a fermion, $\psi$, charged under the three $U(1)$'s in the following way
\begin{equation}
\Big(\partial_{\mu} - i \lambda\ \frac{A_1 - A_2 - A_3}{2}\Big) \psi_{1}
\end{equation}
where
\begin{equation}
A_i = \frac{\cal{B}}{2}(x^k \hat{j} - x^j \hat{k})
\end{equation}
The covariant derivative then becomes
\begin{align}
{D_{\mu}} & = \partial_{\mu} - i\frac{\lambda \cal{B}}{4} \Big\{ (-z + y)\hat{i} + (-x - z) \hat{j} + (y + x) \hat{k} \Big\} \\
&= \left( \begin{array}{ccc} \partial_t  \\  \partial_x \\ \partial_y \\ \partial_z
\end{array} \right) - i\frac{\lambda \cal{B}}{4}\left( \begin{array}{ccc} 0 \\  -z + y \\  -x - z \\ y + x\end{array} \right)
\end{align}
Under a simple change of basis, we can write this in a much more transparent form
\begin{equation}
D_{\mu}'= M_{\mu \nu}D_{\nu} = \left( \begin{array}{ccc} \partial_t  \\  \partial_p \\ \partial_n \\ \partial_m
\end{array} \right) - i\frac{\lambda \cal{B}}{4}\left( \begin{array}{ccc} 0 \\ \sqrt{3} n \\ -\sqrt{3} p \\ 0 \end{array} \right)
\end{equation}
Thus in this basis, the fermion, $\psi_1$, is charged under the gauge field $A' = \frac{\sqrt{3} {\cal{B}}}{2}(n \hat{p} - p \hat{n})$. The conclusion is then that the entropy for this system will have the same behavior as was found previously; it will vanish as $T \rightarrow 0$. Explicitly, the entropy will be
\begin{align}
S &= \frac{\pi}{3} c_F L_y T, 
\\ c_F &= \sum_{\psi} \frac{1}{2} \frac{g |q_{\psi} {\cal{B}}_R| V_2}{2 \pi} \\
&= \frac{1}{2}(|\alpha| + |\beta_1|+ |\beta_2|+ |\beta_3|)\frac{g |{\cal{B}}_R|V_2}{2 \pi}N^2\\
&= \sqrt{3} \ \frac{g |\lambda {\cal{B}}|V_2}{2 \pi}N^2
\end{align}
\indent We remind ourselves with the result of the previous section where we found the entropy to be finite at zero temperature. Thus we have found a conflict in the behaviors of the systems across the duality. The weakly coupled picture suggests a new mechanism for local quantum criticality (A. Almuhairi and J. Polchinski, work in progress).  A charge in a magnetic field can only move parallel to field lines.  For a composite operator whose constituents move in different directions, the correlator will be local in position space, and exhibit critical behavior only in time.

\section{Stability}

In this section we investigate the stability properties of the two space-times AdS$_3\times$T$^2$ and AdS$_2\times$T$^3$ with respect to perturbations of the unimodular tensor field. We consider the perturbation as $T_{i j} = \delta_{i j} +  t_{i j}$, where $Tr[t_{i j}] = 0$ because $det[T_{i j}] = 1$. As usual, we are looking for the field equations of the perturbations from which we can extract the $m^2$; we use the $T_{i j}$ field equations given in  above in eq. \ref{tmotion}. Also across the duality, we expect $T_{i j}$ to correspond to a dimension 2 operator, and thus its $m^2 = -4g^2$ in the case of no magnetic field. This is already saturating the BF-bound. Thus we hope that the magnetic field stabilizes this and does not tip the mass in the wrong direction. The linearized matrix equation of motion is
 \begin{equation}
 \nabla_\mu \nabla^\mu t_{i j}  + 2g[A^\mu, \partial_\mu t]_{i j} + g^2 [A_\mu, 
 [A^\mu, t]]_{i j} = -4g^2 t_{i j} - \frac{1}{2}(t F^2)_{i j} - \frac{1}{2}(F_{\mu \nu}tF^{\mu \nu})_{i j}
\end{equation}
\noindent These set of equations become diagonal when we construct the linear combination of $t$'s: $t_{{z_i} {z_j}}$ and $t_{z_i \bar{z_j}}$, where $z_1 = 1 + i 2, \ z_2 = 3 + i 4, \ z_3 = 5 + i 6$. Due to the traceless property of $t$, the solutions for $t_{z_i \bar{z_i}}$ mix together and we are forced to pick the combinations $t_{z_1 \bar{z_1}} + t_{z_2 \bar{z_2}} - 2 t_{z_3 \bar{z_3}}$, along with the other two cyclic terms. Carrying out the algebra one finds the following $m^2$. 

For AdS$_3$
\begin{align}
&m^2_{z_i \bar{z_j}} = 4g^2 , \ for  \ i \neq j \\
&m^2_{z_i z_j} = 8g^2n, \ for \ all \ i, j \\
&m^2_{z_1 \bar{z_1} + z_2 \bar{z_2} - 2 z_3 \bar{z_3}} = 4 g^2, \ also \ for \ cyclic \ in \ (1,2,3)
\end{align}

For AdS$_2$
\begin{align}
&m^2_{z_i \bar{z_j}} = m^2_{z_i z_j} = 8g^2  n , \ for \ i\neq j \\
&m^2_{z_i z_i} = 4g^2[(n+ \frac{1}{2})2 \sqrt{2} - 1], \ for\ all \ i\\
&m^2_{z_1 \bar{z_1} + z_2 \bar{z_2} - 2 z_3 \bar{z_3}} = 12 g^2, \ also \ for \ cyclic \ in \ (1,2,3)
\end{align}
where $n$ designates the Landau level excitation. We find that all the $m^2$'s are positive and thus safely above the $m^2_{BF}$ for both space-times. Hence the magnetic field acts as a stabilizing mechanism in this problem.

\section{Conclusion}

In this paper we have generalized solutions pertaining to magnetic branes in AdS. We have shown that a more general configuration of magnetic fields gives us control over tuning the ratio between the entropies across the duality in the AdS$_3$ case. There are solutions that can fix the entropies to be identical. This manipulation shows how dependent the physics at large coupling is on the configuration of the magnetic field. \\ 
\indent We also investigated the AdS$_2$ case, where we found that the black hole solution was an extremal one. This gave us a finite entropy at zero temperature. Upon comparing this with the field theory, we observe a conflict, since the field theory entropy scales with temperature. This indicates the presence of a phase transition when turning up the coupling. This is a problem worth tackling. \\
\indent Finally, we studied the stability properties of both space-times and showed that they are stable. In fact we find that the usual dimension two tachyon operator gets a boost in its $m^2$ due to the presence of the magnetic field. Thus the magnetic field acts as a stabilizing agent. 

\subsection*{Acknowledgments}

We wish to thank J. Polchinski for invaluable discussion and guidance.


\begin{thebibliography}{99}



\bibitem{Maldacena:1997re}
  J.~M.~Maldacena,
  ``The large N limit of superconformal field theories and supergravity,''
  Adv.\ Theor.\ Math.\ Phys.\  {\bf 2}, 231 (1998)
  [Int.\ J.\ Theor.\ Phys.\  {\bf 38}, 1113 (1999)]
  [arXiv:hep-th/9711200].
  
\bibitem{Hartnoll:2007ai}
  S.~A.~Hartnoll and P.~Kovtun,
  ``Hall conductivity from dyonic black holes,''
  Phys.\ Rev.\  D {\bf 76}, 066001 (2007)
  [arXiv:0704.1160 [hep-th]].
 
\bibitem{Hartnoll:2007ih}
  S.~A.~Hartnoll, P.~K.~Kovtun, M.~Muller and S.~Sachdev,
  ``Theory of the Nernst effect near quantum phase transitions in condensed
  matter, and in dyonic black holes,''
  Phys.\ Rev.\  B {\bf 76}, 144502 (2007)
  [arXiv:0706.3215 [cond-mat.str-el]].
  
  
\bibitem{Hartnoll:2007ip}
  S.~A.~Hartnoll and C.~P.~Herzog,
  ``Ohm's Law at strong coupling: S duality and the cyclotron resonance,''
  Phys.\ Rev.\  D {\bf 76}, 106012 (2007)
  [arXiv:0706.3228 [hep-th]].
  
    
\bibitem{D'Hoker:2009mm}
  E.~D'Hoker and P.~Kraus,
  ``Magnetic Brane Solutions in AdS,''
  JHEP {\bf 0910}, 088 (2009)
  [arXiv:0908.3875 [hep-th]].
  
\bibitem{Son:2009tf}
  D.~T.~Son and P.~Surowka,
  ``Hydrodynamics with Triangle Anomalies,''
  Phys.\ Rev.\ Lett.\  {\bf 103}, 191601 (2009)
  [arXiv:0906.5044 [hep-th]].
  
\bibitem{Strominger:1998yg}
  A.~Strominger,
  JHEP {\bf 9901}, 007 (1999)
  [arXiv:hep-th/9809027].
  
\bibitem{Faulkner:2009wj}
  T.~Faulkner, H.~Liu, J.~McGreevy and D.~Vegh,
  ``Emergent quantum criticality, Fermi surfaces, and AdS2,''
  arXiv:0907.2694 [hep-th].
  
\bibitem{Faulkner:2010tq}
  T.~Faulkner and J.~Polchinski,
  ``Semi-Holographic Fermi Liquids,''
  arXiv:1001.5049 [hep-th].
  
\bibitem{Cvetic:2000nc}
  M.~Cvetic, H.~Lu, C.~N.~Pope, A.~Sadrzadeh and T.~A.~Tran,
  ``Consistent SO(6) reduction of type IIB supergravity on S(5),''
  Nucl.\ Phys.\  B {\bf 586}, 275 (2000)
  [arXiv:hep-th/0003103].
  
\bibitem{Brown:1986nw}
  J.~D.~Brown and M.~Henneaux,
  ``Central Charges in the Canonical Realization of Asymptotic Symmetries: An
  Example from Three-Dimensional Gravity,''
  Commun.\ Math.\ Phys.\  {\bf 104}, 207 (1986).
  
  
  
  
  \end{thebibliography}
\end{document}